\documentstyle[12pt]{article}

\title{\hfill \\
\vspace{2cm}
Boson Star Rotation: A Newtonian Approximation}
\author{Vanda Silveira\thanks{Email  vanda@guarany.cpd.unb.br} \\
Departamento de Fisica, Universidade de Brasilia\\70910-900 Brasilia, D.F.
Brazil\\
and\\
Claudio M. G. de Sousa\thanks{Email claudio@guarany.cpd.unb.br}\\
Departamento de
Fisica, Universidade de Brasilia\\ 70910-900 Brasilia, D.F. Brazil  \,\,\,\,
and\\Centro Internacional de Fisica da Materia Condensada\\
Universidade de Brasilia\\
CxP 04667, 70919-970 Brasilia, D.F.Brazil}

\begin{document}
\setlength{\baselineskip}{7mm}
\maketitle

\newpage

\begin{abstract}
Using the Newtonian approximation, we study rotating compact bosonic objects.
The equations which describe
stationary states with non-zero angular momentum
are constructed and some numerical results are
presented as examples.
Limits on the applicability of the Newtonian approximation are discussed.
\end {abstract}

\newpage

\section{Introduction}
Boson stars are well known gravitionally bound states of complex scalar fields.
These objects were first studied by Ruffini and Bonazolla [1], and, since then,
a large number of papers on this subject have been published [2], including
three recent reviews [3]. As pointed out by Ferrell and Gleiser [4], we have
two
distinct motivations to study boson stars. On one hand, they make a good
laboratory to study compact self gravitating objects and to explore the
differences and similarities between then and the usual stars. On the other
hand, considering that scalar fields play an important role in theories of
fundamental forces, it is reasonable to address the question of formation and
stability of
bosonic compact objects, including also the study of mixed boson-fermion
objects [5], which may be even more probable to find in nature.

Concerning the analysis of stable solutions of boson stars, most of the work
done is restricted to spherically symmetric configurations, including the
ground state and excitations which are also spherically symmetric. Among the
exceptions, there is the work of Ferrell and Gleiser [4], which analyses the
emission of gravitational radiation by boson stars in the Newtonian
approximation. The compact object is supposed to decay from an excited state
with non zero angular momentum to the ground state with the emission of
gravitational waves. The excited state is treated as a pertubation in which
the excited bosonic particles, carrying the angular momentum, move in the
background potential of the spherically symmetric state.

Recently, Kobayashi, Kasai and Futamase [6] consider
the slow rotation of a relativistic bosonic
star. They look for slowly rotating  solutions which are similiar to the
rotation of conventional objects, following the approach of Hartle [7].
They conclude with a negative result, namely that the relativistic boson
star has no  stationary solutions with slow rotation.
However, the possibility
of rapid rotations, which can not be treated pertubatively, is not excluded.

The failure to describe rotational states pertubatively is itself instructive.
First order pertubation theory is well suited to describe pertubative states
which can be obtained by continuos deformations of the ground state and,
to keep it first order, we must also require that the pertubation parameter is
"small" in some sense.
If the excited rotational states we are searching for form a discrete set,
they can not be obtained by pertubation theory, even if they do not rotate
at relativistic speeds.
And up to now, we do not known whether this compact objects always rotate at
relativistic speeds or not.

In this paper, we return to the problem of rotating boson stars and,
in an attempt
to avoid the pertubative approach, we consider
the Newtonian approximation, which is known to be valid for boson stars
provided that the central density or the total mass of the star is not higher
than a certain critical limit [4].
We assume the hypothesis that, at least for the first excited
states, the rotational effect may be well described by the non-relativistic
theory. This is analogous to the excited states of atoms which are not
obtained as pertubations of the ground state but, even so,  are
well described by the non-relativistic theory.
Under this assumption, the boson star is analysed, and we search for
 stationary solutions with non-zero angular momentum, following the
approach of [4], with one major difference: the excited sates of the scalar
field and their gravitational potential are computed simultaneously in a
coupled system of equations, without any reference to a fixed spherically
symmetric background. The resulting solutions describe
stationary rotational states
which are not pertubative deformations of the ground state.
At the end, we check the validity of the Newtonian approximation and we
identify the limits which justify the
use of  the non-relativistic approximation.

\section{Newtonian Approximation for Boson Stars}
We consider complex scalar fields which are coupled to gravity only. The
action is given by:

\begin{equation}
S= \int \, d^{4}x \,\,\, ({ \frac{R}{16 \pi G} +
g^{\mu \nu} \partial_{\mu} \Phi^{\ast} \partial_{\nu} \Phi -
M^{2} \Phi^{\ast} \Phi })
\end{equation}
Since we will consider only the weak gravity limit of general relativity,
the metric is expanded as \,\,\,
$ g_{\mu \nu} = \eta_{\mu \nu} + h_{\mu \nu}$ \,\,\, with \,\,\,
$|h_{\mu \nu}| \ll 1$ \,\,\, and \\
$\eta_{\mu \nu} = diag( 1, -1, -r^{2}, -r{^2} \sin^{2}(\theta))$.
Using the weak field approximation of General  Relativity [8], we get

\begin{equation}
\Box h_{\mu \nu} = - 16 \pi G S_{\mu \nu}
\end{equation}
where $S_{\mu \nu} = T_{\mu \nu} - \frac{1}{2} \eta_{\mu \nu} T$ and

\begin{equation}
T_{\mu \nu}= \partial_{\mu} \Phi^{\ast} \partial_{\nu} \Phi +
             \partial_{\nu} \Phi^{\ast} \partial_{\mu} \Phi -
 \eta_{\mu \nu}(
 \eta^{\alpha \beta} \partial_{\alpha} \Phi^{\ast}
                     \partial_{\beta} \Phi - M^{2} \Phi^{\ast} \Phi)
\end{equation}
The equation for $\Phi$, derived from (1), is:

\begin{equation}
\Box \Phi + M^{2} \Phi = 0
\end{equation}

We will search for solutions with stationary rotation  so
$\Phi$ will be allowed to depend on $t$ and $\varphi$ only through
a phase, $\Phi(\vec{r}, t) = \phi(r, \theta) e^{i w t} e^{i m \varphi}$.
 $T_{\mu \nu}$, given by (3), will be independent on both $t$ and
$\varphi$, and so will be the metric. To take into account deformations
produced by the rotation, we allow both $\phi$ and $h_{\mu \nu}$ to depend
on $r$ and $\theta$.

Finally, we will restrict ourselves to cases in which special relativistic
effects are not important. The constraint imposed by this restriction
will be checked to be consistent later on this paper. In this approximation,
the only relevant component of $h_{\mu \nu}$ is
$h_{0 0}= 2 V(r, \theta)$, where $V(r, \theta)$ is the Newtonian potential.
Equations (2) and (4) become :

\begin{equation}
\vec{\nabla}^{2} V = 8 \pi G M^{2} \phi^{2}
\end{equation}

\begin{equation}
-D^{2} \phi - (w^{2} - M^{2} )\phi + 2 w^{2} V \phi = 0
\end{equation}
where
$D^{2} \phi = (\vec{\nabla}^{2} - \frac{ m^{2}}{r^{2} \sin^{2} \theta})\phi$,
 $V=V(r, \theta)$ and $\phi= \phi(r, \theta)$.
In the non-relativistic limit, the gravitational binding energy $E$ per
particle
must be much smaller than $M$, and the scalar field frequency may be written
 as $ w= E+M$ with $ |E| \ll M$. The scalar field equation reduces to a
Schrodinger equation:

\begin{equation}
-\frac{1}{2 M} \vec{\nabla}^{2} \phi + M V \phi = E \phi
\end{equation}

Equations (5) and (7) must be solved subjected to the charge
conservation constraint. The gauge invariance of the complex scalar field
implies the conservation of
$j^{\mu} =
i (\partial^{\mu}\phi \phi^{\ast} - \phi \partial^{\mu} \phi^{\ast})$ with
conserved particle number:

\begin{equation}
N= 2 M \int \, \phi^{2} r^{2} \, dr \, \sin(\theta) \, d\theta \, d\varphi
\end{equation}

\section{Stationary Solutions}
We now look for solutions of (5), (7) with non-zero angular momentum described
by $\Phi = e^{i w t} e^{i m \varphi} \phi(r, \theta)$ and $V=V(r, \theta)$.
We expand $\phi(r, \theta)$ in associated Legendre functions:

\begin{equation}
\phi(r, \theta)= \frac{1}{\sqrt{4 \pi}}
\sum_{l=m}^{\infty} R_{l}(r) P^{m}_{l}(\theta)
\end{equation}
and, since $V$, differently from $\Phi$, has no $\varphi$ dependence, we
consider:

\begin{equation}
V(r, \theta) = \sum_{l=0}^{\infty} V_{l}(r) P_{l}(\theta)
\end{equation}
With (9),(10) and the orthogonality relations of $P^{m}_{l}$, we rewrite
(5),(7) as a larger system of equations, which contains one equation for
each value of $l$:

\begin{equation}
V_{l_{0}}'' +  \frac{2}{r}V_{l_{0}}'-\frac{l_{0}(l_{0} +1)}{r^{2}} V_{l_{0}} =
G M^{2} (2 l_{0} +1) \sum_{l, l'} A_{l\,l'\,l_{0}} R_{l} R_{l'}
\end{equation}

\[
\frac{1}{2 M} \left[
 R_{l_{0}}'' + \frac{2}{r} R_{l_{0}}' -
\frac{l_{0}(l_{0} + 1)}{r^{2}} R_{l_{0}} \right] + E R_{l_{0}} =
\]

\begin{equation}
M \frac{(2 l_{0}+1)}{2} \frac{(l_{0} - m)!}{(l_{0} + m)!}
\sum_{l=m}^{\infty} \sum_{l'=m}^{\infty} A_{l\, l_{0}\, l'} R_{l} V_{l'}
\end{equation}
with \, $'=\partial_{r}$ \, and

\begin{equation}
A_{l \, l' \, l_{0}} =
\int^{1}_{-1} \, dx \, P^{m}_{l}(x) \, P^{m}_{l'}(x) \, P_{l_{0}}(x)
\end{equation}

This system of equations may be rescaled by introducing the new variables
[4]:

\[
\hat{r}= \hat{N} M r,
\]

\[
V(r, \theta) = \hat{N}^{2} \hat{V}(\hat{r}, \theta),
\]

\[
R(r) = \hat{N}^{2} (2 G)^{-1/2} \hat{R}(\hat{r}),
\]

\[
E= M \hat{N}^{2} \hat{E}
\]
and
\begin{equation}
\hat{N}= G M^{2} N \frac{(2 l+1)(l-m)!}{(l+m)!}
\end{equation}
With this new variables, our system can be written as:

\begin{equation}
\hat{V}_{l_{0}}'' +  \frac{2}{\hat{r}}\hat{V}_{l_{0}}'-\frac{l_{0}(l_{0}
+1)}{\hat{r}^{2}}\hat{ V}_{l_{0}} =
\frac{(2 l_{0} +1)}{2}
\sum_{l, l'} A_{l\,l'\,l_{0}} \hat{R}_{l} \hat{R}_{l'}
\end{equation}

\[
\frac{1}{2} \left[
\hat{ R}_{l_{0}}'' + \frac{2}{\hat{r}}\hat{ R}_{l_{0}}' -
\frac{l_{0}(l_{0} + 1)}{\hat{r}^{2}}\hat{ R}_{l_{0}} \right] +
\hat{ E} \hat{ R}_{l_{0}} =
\]

\begin{equation}
 \frac{(2 l_{0}+1)}{2} \frac{(l_{0} - m)!}{(l_{0} + m)!}
\sum_{l=m}^{\infty} \sum_{l'=m}^{\infty} A_{l\, l_{0}\, l'}\hat{ R}_{l}
\hat{V}_{l'}
\end{equation}
where $'$ now stands for $\frac{\partial}{\partial \hat{r}}$ and we must add
the normalization condition derived from (8):

\begin{equation}
\int \hat{R}_{l_{0}}^{2}(\hat{r}) \,\, \hat{r}^2 \,\, d\hat{r} = 1
\end{equation}

We may now systematically search for solutions of (15), (16) for different
values of $l$ and $m$. The basic idea is to consider solutions with
$\hat{R}_{L} \neq 0$ for one particular value of $l$, $L$, with
$\hat{R}_{l}$ for  $l \neq L$. The ground state equations correspond to
the choice   $m=0$ with  $\hat{R}_{l} = 0$ for $l \neq 0$, in which case the
equations simplify to:

\[
\hat{V}_{l}'' + \frac{2}{\hat{r}} \hat{V}_{l}' =
\frac{1}{2} A_{0 \, 0 \, l} (\hat{R}_{0})^{2}
\]

\begin{equation}
\frac{1}{2} \left[
\hat{ R}_{0}'' + \frac{2}{\hat{r}}\hat{ R}_{0}' \right] +
\hat{ E} \hat{ R}_{0} =
 \frac{1}{2}
 \sum_{l'=0}^{\infty} A_{0\, 0\, l'}\hat{ R}_{0} \hat{V}_{l'}
\end{equation}
and $A_{0 \, 0 \, l}= \int_{-1}^{1} \, dx \,\, P_{l}(x) = 2 \delta_{l \, 0}$.
For $l \neq 0$, the equation for $\hat{V}_{l}$ is homogeneous and the
solution is the trivial $\hat{V}_{l}(\hat{r})=0$. In the expansion (10), the
only
non-zero component is $\hat{V}_{0}$ and we are left with the simple system:

\[
\hat{V}_{0}'' + \frac{2}{\hat{r}} \hat{V}_{0}' =
 (\hat{R}_{0})^{2}
\]

\begin{equation}
\frac{1}{2} \left[
\hat{ R}_{0}'' + \frac{2}{\hat{r}}\hat{ R}_{0}' \right] +
\hat{ E} \hat{ R}_{0} =
 \hat{ R}_{0} \hat{V}_{0}
\end{equation}

The excited states will correspond to other choices of $\hat{R}_{L} \neq 0$.
As an example, we may focus on the $l=2$, $m=0$ state described by the
system of equations (15), (16) with the appropriate values of $l$ and $m$.
We set  $\hat{R}_{l}=0$ for $l \neq 2$.
The r.h.s of equations (15) are given by
$\frac{2 l_{0} + 1}{2} A_{2 \, 2 \, l_{0}} (\hat{R}_{2}(\hat{r}))^{2}$ where
$A_{2 \, 2 \, l_{0}} = \int_{-1}^{1} \, dx \, (P_{2}(x))^{2} \, P_{l_{0}}(x)$,
with the following numerical values: $A_{2 \, 2 \, 0} = 2/5$,
$A_{2 \, 2 \, 2} = 4/35$, $A_{2 \, 2 \, 4}= 4/35$,
$A_{2 \, 2 \, l} = 0$ for $l > 4$ and for odd values of $l$.
So, for $l=2$ and $m=0$, the Newtonian potential will be given by:

\begin{equation}
V(r, \theta)= V_{0}(r) + V_{2}(r) \, P_{2}(\theta) + V_{4}(r) \, P_{4}(\theta)
\end{equation}
For $l>4$, $V_{l}$, as solution of an homogeneuos system, may be set equal
to zero. $V_{0}$, $V_{2}$, $V_{4}$ and $R_{2}$ will be given as solutions
of:

\begin{equation}
\hat{V}_{0}'' + \frac{2}{\hat{r}} \hat{V}_{0}' =
\frac{1}{5} (\hat{R}_{2})^{2}
\end{equation}

\begin{equation}
\hat{V}_{2}'' + \frac{2}{\hat{r}} \hat{V}_{2}' -
\frac{6}{\hat{r}^{2}} \hat{V}_{2} =
\frac{2}{7} (\hat{R}_{2})^{2}
\end{equation}

\begin{equation}
\hat{V}_{4}'' + \frac{2}{\hat{r}} \hat{V}_{4}' -
\frac{20}{\hat{r}^{2}} \hat{V}_{4} =
\frac{18}{35} (\hat{R}_{2})^{2}
\end{equation}

\begin{equation}
\frac{1}{2} \left[
\hat{ R}_{2}'' + \frac{2}{\hat{r}}\hat{ R}_{2}' -
               \frac{6}{\hat{r}^2}\hat{R}_{2}   \right]   +
\hat{ E} \hat{ R}_{2} =
 \hat{ R}_{2} \left[ \hat{V}_{0} +\frac{2}{7} \hat{V}_{2} +
                                  \frac{2}{7} \hat{V}_{4} \right]
\end{equation}
with the appropriate boundary conditions. For $V_{0}$, $V_{2}$ and $V_{4}$,
we may incorporate these boundary conditions by formally solving (21),(22),(23)
with the help of Green's functions. As usual, we consider the solution of:

\begin{equation}
G_{l}''(r, r') + \frac{2}{r} G_{l}'(r, r') -
\frac{l(l+1)}{r^{2}} G_{l}(r, r') =
\frac{1}{r^{2}} \delta (r- r')
\end{equation}
given by :

\begin{equation}
G_{l}(r,r') = -\frac{1}{(2 l +1)} \frac{r_{<}^{l}}{r_{>}^{l+1}}
\end{equation}
which is regular at $r \rightarrow 0$ and goes to zero for
$r \rightarrow \infty$.

In the particular case we are considering, namely, $l=2$ and $m=0$,
we end up with:

\begin{equation}
\hat{V}_{0}(r) = - \frac{1}{5} \left[
\frac{1}{\hat{r}} \int_{0}^{\hat{r}} \, dr'\, (r')^{2}\, (\hat{R}_{2}(r'))^{2}+
                  \int_{\hat{r}}^{\infty}\, dr'\, r' (\hat{R}_{2}(r'))^{2}
                  \right]
\end{equation}

\begin{equation}
\hat{V}_{2}(r) = - \frac{2}{35} \left[
\frac{1}{\hat{r}^{3}} \int_{0}^{\hat{r}} \, dr'\, (r')^{4}\,
(\hat{R}_{2}(r'))^{2}
+ \hat{r}^{2}  \int_{\hat{r}}^{\infty}\, dr'\, \frac{1}{r'}\,
(\hat{R}_{2}(r'))^{2}
                  \right]
\end{equation}

\begin{equation}
\hat{V}_{4}(r) = - \frac{2}{35} \left[
\frac{1}{\hat{r}^{5}} \int_{0}^{\hat{r}} \, dr'\, (r')^{6}\,
(\hat{R}_{2}(r'))^{2}
+ \hat{r}^{4}  \int_{\hat{r}}^{\infty}\, dr'\, \frac{1}{(r')^{3}}\,
(\hat{R}_{2}(r'))^{2}
                  \right]
\end{equation}
and we are left with just one (integral-differential) eigenvalue equation
for $\hat{R}_{2}(\hat{r})$:

\begin{equation}
\frac{1}{2} \left[
\hat{ R}_{2}'' + \frac{2}{\hat{r}}\hat{ R}_{2}' -
               \frac{6}{\hat{r}^2}\hat{R}_{2}   \right]
- \hat{R}_{2}  W(\hat{r}) =
- \hat{ E_{2}} \hat{ R}_{2}
\end{equation}
with

\begin{equation}
W(\hat{r}) = \hat{V}_{0}(\hat{r}) +
       \frac{2}{7} \hat{V}_{2}(\hat{r}) +
       \frac{2}{7} \hat{V}_{4}(\hat{r}) \,,
\end{equation}

$\int_{0}^{\infty} \, d\hat{r} \, \hat{r}^{2} \, (\hat{R}_{2}(\hat{r}))^{2}=1$,
and $\hat{R}_{2}(0) =0$.

The procedure described in the $l=2$, $m=0$ example may be applied to all
excited states. We should note here that the infinite sums in
the r.h.s. of (15), (16) will
always reduce to  finite sums because
$\left[ P_{l_{0}}^{m}(x) \right]^{2}$ is a polynomial of
order $l^{2}_{0}$, and may be expanded as a sum of Legendre polynomials
with $l \leq 2 \, l_{0}$.
So, using the orthogonality of the Legendre polynomials,
 for any fixed value of $l_{0}$,
 $A_{l_{0} \, l_{0} \, l} =0$ for $l> 2 \, l_{0}$.

Without any intention to make a complete  analysis of these
excited states, we present the numerical results for $l=0$, $m=0$, the
groundstate, and for the excited states $l=1$, $m=0$ ; $l=1$, $m=1$ ;
$l=2$, $m=0$. Basically, we start with a trial function,
$R^{i=0}(\hat{r})$, satisfying the appropriate boundary conditions.
The index $i$ gives the iteration order.
We construct $W(\hat{r})$ using (31), or the equivalent expression for other
states, and we use it to obtain a new $R^{i=1}(\hat{r})$, with the
corresponding
eigenvalue $E^{i=1}$. The process is then repeated with $R^{i=1}$, and so on,
until convergence is achieved in the solutions and, up to a given precision,
no effect is introduced by new iterations. Our results are plotted in
the figures 1, 2, 3. Note that the excited states are
qualitatively different from the ground state and this is why they can not be
obtained as simple pertubations
of the ground state.

\section{Conclusion}
Following the prescription outlined here,
in principle, we may construct all the excited levels of
the rotating boson star, and based on these states, predictions can
be made about the spectrum produced by boson  stars decaying from excited
states
to states with lower energy.
However, we still must check the consistence of the non-relativistic
approximation made at the beginning.

Basically, we must require that the gravitational field is weak and
that no relativistic speeds are present. The weakness of the field
is verified by requiring that the
mass of the star in the Newtonian approximation
is only slightly affected by rotation.
The total mass of each state, $M_{T}$,  is given by [4]:

\begin{equation}
M_{T}= M \, N + N \, E =
\hat{N}(\frac{M^{2}_{Planck}}{M})(1 + \hat{E} \hat{N}^{2})
\end{equation}
The rotation of the star changes its mass by the relative amount:

\begin{equation}
\Delta=  \frac{M_{T}^{excited}- M_{T}^{0}}{M_{T}^{0}} =
\frac{(\hat{E}_{excited} \hat{N}^{2}-\hat{E}_{0} \hat{N}^{2}_{0})}
{1+\hat{E}_{0} \hat{N}^2_{0}}
\end{equation}
where we are comparing states with the same total number $N$ of particles,
and $N$ is related to $\hat{N}_{0}$ and $\hat{N}$ by (14).
To be consistent with the Newtonian approximation, we should only apply
the above results for boson stars with $\Delta \ll 1$.
Since $|\hat{E}_{0}| \sim 10^{-1}$ and
$|\hat{E}_{excited}| < |\hat{E}_{0}|$,
$\Delta \ll 1$ is always satisfied by
$\hat{N}^{2}_{0} \ll 1$.
For $\hat{N}^{2}_{0} \sim 10^{-2}$, the boson star has a mass of order
$10^{10} \, kg$, too small to resemble a conventional stellar object.
However, this number is not much different from the maximum
mass of boson stars
 of free scalars in the ground state, which is roughly $10^{11} \, kg$.

The restriction coming from the small velocities limit is more severe.
To check for relativistic speeds, we assume that a global effective
angular velocity $\Omega^{eff}$ may be associated to the stationary
excited state, with:

\begin{equation}
L^{eff}_{z} = I\,\, \Omega^{eff} \,\,\,\,\, and \,\,\,I=\int \, \rho \,z^2\, dV
\end{equation}
and $\rho$ is the energy density given by $T_{0 0}$.
Taking into account that $I>I_{0}$ where $I_{0}$ is the moment of inertia of
the ground state configuration, we have:

\begin{equation}
L^{eff}_{z} > I_{0}\,\, \Omega^{eff} =
\frac{2}{3} <r^{2}_{0}> M_{0}^{T}\,\, \Omega^{eff}
\end{equation}
with $<r^{2}_{0}> = \frac{\int \rho_{0} r^{2} dV}{\int \rho_{0} dV}=
                    \frac{1}{M_{0}^{T}} \int \rho_{0} r^2 dV$.

This effective angular momentum is then identified to the angular
momentum of the configuration obtained by direct integration of
 the angular momentum tensor:

\begin{equation}
L_{z} = \int \,(x T_{0y} - y T_{ox}) = m w \frac{N}{M}
\end{equation}
Taking $m=l$ and comparing (35) and (36), we have:

\begin{equation}
\Omega^{eff}\,< \frac{3 l }{2 M <r^{2}_{0}>} \frac{w}{w_{0}}
\end{equation}
and $v^{max} \sim 2 <r> \Omega^{eff}$ satisfies :

\begin{equation}
v^{max} < 3 \,\, \frac{<\hat{r}>}{<\hat{r}^{2}_{0}>}\,\,\frac{w}{w_{0}}\,\,
\hat{N}_{0}\,\, \frac{l (l+m)!}{(2 l+1)(l-m)!}
\end{equation}
A sufficient condition to guarantee the use of non-relativistic
approximation is:

\begin{equation}
 3 \,\, \frac{<\hat{r}>}{<\hat{r}^{2}_{0}>}\,\,\frac{w}{w_{0}}
\,\,\hat{N}_{0} \,\,\frac{l (l+m)!}{(2 l+1)(l-m)!} \ll 1.
\end{equation}
Now, $w \sim w_{0}$, but $<\hat{r}>$ grows fast with the excitation
levels and even faster grows the factor
$\frac{l(l+m)!}{(2 l+1)(l-m)!}$.

Our conclusion is that the non-relativistic
rotation is possible only for very small bosonic objects (small values for
$\hat{N}_{0}$) and, even starting with small $\hat{N}_{0}$, only the first
excited states will be well describe by the Newtonian theory.
For higher excited states and objects with $\hat{N}_{0} \sim 1$,
 relativistic effects will be important and we can rely only
on a fully relativistic calculation.
These restrictive limits tell us that the numerical results of our calculation
can not be applied to large, astrophysically important stellar objects.
Even so, we showed that small compact bosonic objects  rotate producing
a discrete set of excited states and we set limits on the
applicability of the Newtonian approximation.
Since the introduction of self-coupling is known to increase the maximum mass,
it would be interesting to study how the self-coupling changes the rotational
states, a problem we hope to address in the future.

\section{Acknowledgments}
The authors would like to thank the  CNPq for the finantial support.

\newpage

Figure Captions\\
\vspace{15mm}

Fig. 1a. Ground state ($l=0$, $m=0$) function, $\hat{R}(\hat{r})$, normalized
to $\int_{0}^{\infty} \, d\hat{r} \, \hat{r}^{2}\, (\hat{R}(\hat{r}))^{2}=1$,
for the last twelve iterations. $\hat{E}_{l=0,m=0}= - 0.16$.

Fig. 1b. Newtonian potential for each iteration in the same case
($l=0$, $m=0$). \\
\vspace{8mm}

Fig. 2. Excited states also normalized to
$\int_{0}^{\infty} \, d\hat{r} \, \hat{r}^{2}\, (\hat{R}(\hat{r}))^{2}=1$.
Curve 1: $l=1, \,\, m=1, \,\, \hat{E}_{1,1}= - 0.025$ ;
Curve 2: $l=1, \,\, m=0, \,\, \hat{E}_{1,0}= - 0.0071 $;
Curve 3: $l=2, \,\, m=0, \,\, \hat{E}_{2,0}= - 0.0013$. \\
\vspace{8mm}

Fig. 3. Newtonian potential for $l=2, \,\,m=0$, showing contributions from
$\hat{V}_{0}(\hat{r})$, $\hat{V}_{2}(\hat{r})$ and $\hat{V}_{4}(\hat{r})$.
The potential well, with constant, non-zero value for $r \rightarrow 0$ is
$\hat{V}_{0}(\hat{r})$. The other two represent $\hat{V}_{2}(\hat{r})$ and
$\hat{V}_{4}(\hat{r})$.

\end{document}